\documentclass[12pt]{article}

\usepackage{amsmath,amsfonts,amssymb,latexsym,graphicx}

\newtheorem{theorem}{Theorem}[section]

\numberwithin{equation}{section}
\def\be{\begin{equation}}
\def\ee{\end{equation}}
\def\bq{\begin{eqnarray}}
\def\eq{\end{eqnarray}}
\def\beq{\begin{eqnarray}}
\def\eeq{\end{eqnarray}}

\def\a{\alpha}

\begin{document}

\title{\Huge{\textsc{The singularity problem in brane cosmology}}}
\author{{\Large\textsc{Ignatios Antoniadis$^{1,2}$\thanks{\texttt{antoniad@lpthe.jussieu.fr}}, Spiros Cotsakis\footnote{On leave from the University of the Aegean, Greece.}\,\,$^{3}$\thanks{\texttt{skot@aegean.gr}}}}, 
\\[10pt]
$^1$LPTHE, UMR CNRS 7589, Sorbonne Universit\'es, UPMC Paris 6,\\
4 place Jussieu, T13-14, 75005 Paris, France\\
$^2$ Albert Einstein Center for Fundamental Physics, ITP,\\
University of Bern, 
Sidlerstrasse 5 CH-3012 Bern, Switzerland\\
$^{3}$Department of Mathematics, 
American University of the Middle East\\
P. O. Box 220 Dasman, 15453, Kuwait }

\maketitle
\begin{abstract}
\noindent
We review results about the development and asymptotic nature of singularities in `brane-bulk' systems. These arise  for  warped metrics obeying the 5-dimensional Einstein equations with fluid-like sources, and including a brane 4-metric that is either Minkowski or de Sitter or Anti-de Sitter. We characterize all singular Minkowski brane solutions, and look for regular solutions with nonzero curvature. We briefly comment on matching solutions, energy conditions and finite Planck mass criteria for admissibility, and we  briefly discuss the connection of these results to ambient theory.

\end{abstract}
\section{Introduction}
The singularity problem in the setup of brane cosmological models is concerned with the existence and nature of the dynamical singularities that may arise when one considers the evolution of metrics and fields propagating in spaces with `large extra dimensions' containing certain lower-dimensional slices. Such systems obey higher-dimensional  Einstein (or possibly similar string gravity) equations, with the standard interactions usually confined in a 4-dimensional slice (the brane) sitting in a 5-(or higher-)dimensional spacetime (the bulk). Such `brane-bulk' systems are used in an essential way as a means to overcome the hierarchy problem \cite{arc,ign}, and in a crucial way in approaches to solve the cosmological constant problem \cite{nima,silver}.

In this paper we provide a concise overview of the various ramifications and results that have been obtained in recent years about the singularity problem in such contexts, basically using the methods developed in Refs. \cite{ack1}-\cite{ack4}. Previous work on this subject can be found in \cite{gubser}. We also briefly discuss the connection of these results with the ambient approach to the singularity problem towards the end of this work.

We write the \emph{bulk metric} in the form,
\be
\label{warpmetric}
g_{5}=a^{2}(Y)g_{4}+dY^{2},
\ee
where $g_{4}$ represents the \emph{brane metric}, taken to be either Minkowski, or  de Sitter (dS) or anti-de Sitter (AdS),
\be
\label{branemetrics}
g_{4}=-dt^{2}+f^{2}_{k}g_{3},
\ee
with
\be
\label{g_3}
g_{3}=dr^{2}+h^{2}_{k}g_{2},\quad g_{2}=d\theta^{2}+\sin^{2}\theta d\varphi^{2},
\ee
where $f_{k}=\cosh (H t)/H$ or $\cos (H t)/H $ ($H^{-1}$ is the de Sitter (or AdS)
curvature radius) and $ h_{k}=\sin r$ or $\sinh r $, for dS or AdS respectively.

There are two interesting interpretations of the metric  (\ref{warpmetric}) that  are relevant in the present context. The first is of course the standard one, namely, to view the brane as a domain wall solution, a hypersurface in the 5-space, the bulk. This is the most common interpretation of the geometric setup, across the entire braneworld literature (cf. \cite{gasp} and references therein). There is, however, a different one, useful in certain contexts (cf. especially the discussion towards the end of the present paper), namely, to view the metric (\ref{warpmetric}) as a cone metric, or a warped product metric \cite{peterson,o'neill}.

Whatever the geometric interpretation, we impose the 5-dimensional Einstein equations on the metric (\ref{warpmetric}),
\be\label{ei}
G_{AB}=\kappa^{2}_{5}T_{AB},
\ee
where we shall usually  take the energy-momentum tensor to be that of an analog of a 5-dimensional (5d) fluid (with the Y coordinate playing the role of time), or a combination of fluids, possibly exchanging energy. In fact, it is an interesting result that our 5d-fluid must by necessity be an anisotropic pressure fluid. (Such fluids have recently emerged as important instability factors in other contexts in string cosmology, for example in the possible disruption of the isotropic fluid stability of simple ekpyrotic cyclic models \cite{by}.) To see this, we start with the standard energy-momentum tensor for the 5d-fluid in the form,
\be
\label{T old}
T_{AB}=(\rho+P)u_{A}u_{B}-Pg_{AB},
\ee
where $A,B=1,2,3,4,5$ and $u_{A}=(0,0,0,0,1)$, with the 5th coordinate corresponding to $Y$, and seek an anisotropic pressure form,
\be
\label{T new}
T_{AB}= (\rho^{0}+
p^{0})u_{A}^{0}u_{B}^{0}
+p^{0}g_{\alpha\beta}\delta_{A}^{\alpha}\delta_{B}^{\beta}+
p_{Y}g_{55}\delta_{A}^{5}\delta_{B}^{5},
\ee
where $u_{A}^{0}=(a(Y),0,0,0,0)$ and $\alpha,\beta=1,2,3,4$.
When we combine (\ref{T old}) with (\ref{T new}), we find that the 5d-fluid has an anisotropic energy-momentum tensor of the form  \cite{ack3,ack4},
\be
T_{AB}= -P g_{\alpha\beta}\delta_{A}^{\alpha}\delta_{B}^{\beta}+
\frac{P}{\gamma}g_{55}\delta_{A}^{5}\delta_{B}^{5},
\ee
when $P=\gamma \rho$. We see that isotropic fluids in this context correspond to the limiting case of a cosmological constant-like equation of state, $\gamma\rightarrow -1$. We can then satisfy the various energy conditions by
restricting $\gamma$ to take values in certain intervals \cite{ack3,ack4}.

It is important to further point out that in this work, except for a fixation of the braneworld 4-geometry (either Minkowski, or dS or AdS, respecting 4d maximal symmetry), we do \emph{not} fix the bulk 5-geometry other than take it to be of the above warped type near the (presumed) singularity. Hence, only the \emph{asymptotic} geometry of the bulk is found and dictated by the 5-dimensional Einstein equations with the fluid source discussed above. Away from such an open neighborhood around the singularity, the bulk space geometry remains compatible with that requirement. This is in sharp contrast with other approaches, e.g., \cite{rs1}, where the bulk is fixed rigidly to be of some preassigned form (eg., $AdS_5$).

\section{Flat branes}
The prototype case for the evolution of the brane-bulk system near its finite-distance singularities is when the brane is described by Minkowski space and there is a single, free scalar field $\phi$ in the bulk. In this case,  the 5-dimensional Einstein equations (\ref{ei}) in the bulk with source $\phi$, can be symbolically written as an autonomous dynamical system in the form,
\be\label{DS}
\dot{X}=f(X).
\ee
All solutions of this system then become the integral curves of the 3-dimensional, non-polynomial vector field \cite{ack1},
\be
f(X)=\left(y,-\lambda Az^{2}x,-4yz/x\right),
\ee
subject to the constraint,
\be
\label{constraint_flat} \frac{y^{2}}{x^{2}}=\frac{A\lambda}{3}
z^{2}.
\ee
Here, $X=(x,y,z)$, $A,\lambda$ are constrants, while we have introduced new variables by setting
\be
x=a, \quad y=a', \quad z=\phi',
\ee
with a prime denoting differentiation with respect to the extra dimension $Y$. Then we have the following result (\cite{ack1}, Section 2.1).
\begin{theorem}[Minkowski brane-massless dilaton]
With the setup of a flat 3-brane in a 5-dimensional bulk spacetime filled with a free scalar field as described above, let $Y_s$ denote the position of the finite-distance singularity from the brane position. Then there is  only one possible asymptotic behaviour of the solutions of the field equations towards to singularity, given by,
\be\label{as1}
a\rightarrow 0,\quad a'\rightarrow \infty,\quad \phi'\rightarrow \infty,
\ee
 as $Y\rightarrow Y_{s}$.
\end{theorem}
This result means that all solutions asymptote towards a state wherein the flat brane collapses after `traveling' a finite distance in the bulk starting from its initial position, with the energy of the scalar field blowing up there. It implies that any initial configuration involving a Minkowski 3-brane coupled to a bulk massless dilaton satisfying the 5-dimensional Einstein equations, will gradually evolve to the collapse state described in the Theorem above. This result completely fixes the nature of the singularity in this simple case and the behaviour of all solutions near the singularity. (An exact, particular solution with these properties was first found in \cite{nima}, \cite{silver}. One may view the result contained in the Theorem above as implying that the exact solution found in those references is a stable one in the sense that all other solutions of the system approach this form asymptotically towards the singular point.)

However, one naturally wonders whether the above result has some degree of genericity, in other words, whether and how the existence and nature of the singularity in this simplest Minkowski brane model persists when one passes on to more general ones, while keeping the flatness assumption. (The extension to branes with curvature is separately discussed in the next Section of this paper.) There are at least three ways to treat the flat brane problem in a more general setting:
\begin{itemize}
\item Add self-interaction to the dilaton
\item Add a perfect fluid in the bulk
\item Add a mixture of a fluid and a (possibly interacting) dilaton field.
\end{itemize}
When we turn to a Minkowski brane-fluid bulk system instead of a massless dilaton bulk, is that although the existence of the finite-distance singularity remains (except perhaps moved on to the envelope-see below), its \emph{nature} depends on the range of the equation of state fluid parameter $\gamma$ defined by the equation $P=\gamma\rho$. For a Minkowski brane, the Einstein equations (\ref{DS}) give
 \be
\label{vectorfield} \mathbf{f}=\left(y,-2A\frac{(1+2\gamma)}{3}w
x,-4(1+\gamma) \frac{y}{x}w\right),
\ee
subject to the constraint,
\be
\label{constraint_flat2} \frac{y^{2}}{x^{2}}=\frac{w}{\delta},\quad\delta=3/2A,
\ee
with $A=\kappa^2_5/4$, and where for the new variables of this problem we introduce the definitions \be x=a, \quad y=a', \quad w=\rho, \ee with $\rho$ being the fluid energy density. Then we have the following result (\cite{ack1}, Section 3)
\begin{theorem}[Minkowski brane-Single bulk fluid]
With the setup of a flat 3-brane in a 5-dimensional bulk spacetime filled with a fluid  as described above, let $Y_s$ denote the position of the finite-distance singularity. Then the possible asymptotic behaviours of the solutions of the field equations are all singular, have the required number of arbitrary constants to qualify as corresponding to a general solution,  and are  given by,
\begin{itemize}
\item Collapse-type I: $\gamma>-1/2$
\be\label{as2}
a\rightarrow 0,\quad a'\rightarrow \infty,\quad \rho\rightarrow \infty,
\ee
\item Collapse-type II: $\gamma=-1/2$
\be\label{as2}
a\rightarrow 0,\quad a'\rightarrow\textrm{const.},\quad \rho\rightarrow \infty,
\ee
\item Big rip: $\gamma<-1$
\be\label{as3}
a\rightarrow \infty,\quad a'\rightarrow -\infty,\quad \rho\rightarrow \infty,
\ee
\item At envelope: $\gamma\in (-1,-1/2)$
\be\label{as2}
a\rightarrow 0,\quad a'\rightarrow 0,\quad \rho\rightarrow \infty,
\ee
\end{itemize}
 as $Y\rightarrow Y_{s}$.
\end{theorem}
This result implies generally speaking that the situation described by Theorem 2.1 is still valid when we pass to the more general fluid content of Theorem 2.2: Minkowski brane-fluid systems are generically singular and behave basically like the massless dilaton case. For example, item 1 in the Theorem 2.1 means that the runaway situation of Theorem remains valid for any fluid having $\gamma>-1/2$. A slightly milder singularity is approached by the flat brane-fluid systems when $\gamma=-1/2$, the singular point is attained with bounded speed. The approach to the singularity at a finite distance from the brane changes its nature to that of a big rip when the bulk fluid is phantom-like.

The last item in Theorem 2.2 requires a separate, more involved analysis based on the observation that when solutions of a differential equation have an envelope, the dominant balance picks the envelope not the general solution, and therefore instead of looking at enveloping solutions from the general solution, we may proceed to construct such solutions directly from the field equations \cite{ack3}. Eq. (\ref{as2}) then implies that all solutions of the field equations having $\gamma\in (-1,-1/2)$  asymptote to the singular, first component $\Sigma_1$ of an `enveloping brane' defined as a disjoint\footnote{We use the term `disjoint' because their common element, namely, $(0,0,0)$, is not a realizable state asymptotically.} union (cf. \cite{ack3}, Sec. 2, where this result is proved),
\be\label{pfenv}
\Sigma=\Sigma_1\bigsqcup\Sigma_2,
\ee
where
\be\label{pfenv1}
\Sigma_1:\quad x=0,\quad y=\pm H\sqrt{k},
\ee
\be\label{pfenv2}
\Sigma_2:\quad y=\pm H\sqrt{k},\quad w=0.
\ee

The impossibility of regular solutions away from a Minkowski brane with a massless dilaton or single fluid as sources in the bulk, prompts us to search for such solutions further, by considering \emph{mixtures} of the two in the bulk. This is an on-going project with many open problems, the non-interacting, co-existing fluid case treated in detail in Ref.  \cite{ack2}. For the massless scalar we then take an energy-momentum tensor of
the form $T^{1}_{AB}=(\rho_{1}+P_{1})u_{A}u_{B}-P_{1}g_{AB}$
where $A,B=1,2,3,4,5$, $u_{A}=(0,0,0,0,1)$ and $\rho_{1}$, $P_{1}$
are its density and pressure  which we take as
$P_{1}=\rho_{1}=\lambda\phi'^{2}/2$, with $\lambda$ a
parameter. For the second fluid, we assume that $T^{2}_{AB}=(\rho_{2}+P_{2})u_{A}u_{B}-P_{2}g_{AB}$, and an
equation of state of the form $P_{2}=\gamma\rho_{2}$.
Here $\rho_{1}$, $\rho_{2}$ and $P_{1}$, $P_{2}$ are functions of the fifth
dimension $Y$ only. The five-dimensional Einstein field equations (\ref{ei})  in the case of a flat (Minkowski) brane assume a more complicated form, basically a neat problem in bifurcation theory. Namely,
\bq
\label{syst1_1_n_s}
x'&=&y\\
y'&=&-A \lambda z^{2}x-\dfrac{2}{3}A(1+2\gamma)wx\\
z'&=&-\left(4+\dfrac{\nu}{2}\right) \dfrac{y z}{x}+\frac{\sigma}{\lambda}\frac{y w}{x z}\\
\label{syst1_4_n_s}
w'&=&-(4(\gamma+1)+\sigma)\dfrac{y w}{x}+\frac{\lambda \nu}{2}\frac{y z^{2}}{x},
\eq
with the constraint,
\be
\label{constraint} \dfrac{y^{2}}{x^{2}}=\dfrac{A\lambda}{3}z^{2}+
\dfrac{2A}{3}w.
\ee
We write $(x,y,z,w)=(a,a',\phi',\rho_{2})$, and the new system has four parameters $\lambda, \gamma, \sigma, \nu$, the last two describing the possible exchange of energy between the two components, no exchange of energy corresponding to the case $\nu=\sigma=0$. This is the main case that is analyzed in Ref. \cite{ack2}. The main result is this.
\begin{theorem}[Minkowski brane: Non-interacting pair of massless dilaton-fluid]
With the setup  as described above, let $Y_s$ denote the position of the finite-distance singularity. Then the possible asymptotic behaviours of the solutions of the field equations are all singular, have the required number of arbitrary constants to qualify as corresponding to a general solution,  and are  given by,
\begin{itemize}
\item Collapse-type I: any $\gamma$
\be\label{as4}
a\rightarrow 0,\quad  a'\rightarrow \infty,\quad  \phi'\rightarrow \infty,\quad
\rho_{2}\rightarrow 0, \rho_{s}, \infty,
\ee

\item Big rip: $\gamma<-1$
\be\label{as5}
a\rightarrow \infty,\quad a'\rightarrow -\infty,\quad  \phi'\rightarrow 0,\quad
\rho_{2}\rightarrow\infty.
\ee

\end{itemize}
 as $Y\rightarrow Y_{s}$.
\end{theorem}
We observe that in these asymptotic solutions the final states are characterized by the asymptotic dominance of the dilaton over the fluid component. This is the reason why we obtain singularities for all possible asymptotic balances but of a similar character as the massless dilaton case. In particular, there cannot be any stable asymptotic situation wherein the fluid attains some finite asymptotic value with vanishing dilaton. This is reminiscent of the generic early behaviour of scalar-tensor cosmologies where there is a complete dominance of the scalar field over matter.

The inclusion of an interacting pair of dilaton-fluid could in principle lead to regular solutions away from the Minkowski brane. In Ref. \cite{ack2} it was noticed that by suitably choosing the exchange parameters $\nu,\sigma$ and analyzing the resulting dynamical system, has the effect of moving these singular points to infinity. There is an intricate structure of the eigenvalues of the asymptotic matrix that controls the behaviour of the solutions in this case, and this structure leads to the interesting result that for the same interval of the fluid parameter as in the massless dilaton case, namely $\gamma\in(-1,-1/2)$, the singularities are seen to move to infinity. However, we expect that they are just moved to the singular envelope as before, therefore not being true regular solutions. The generic problem, however, is entirely open at present.

\section{Curved branes}
Making the brane positively or negatively curved, has the apparent effect of moving the singularities to infinite distance away from the original brane position. However, this may just mean that we are looking at the enveloping brane. The problem then is to determine the precise extent of the singular and regular parts of the enveloping set. This may be an intricate problem. Below we call a solution \emph{regular} if the scale factor is non-collapsing or divergent in a finite distanace away from the brane. This does not exclude the density from having a singularity at the envelope.

With just a massless dilaton support in the bulk, one indeed may get regular curved brane solutions with a decaying dilaton. However, to get this one has to sacrifice an arbitrary constant, ending up with a family that does not correspond to a general solution of the field equations, at least for de Sitter branes, (for AdS branes there may be no such restriction) cf. \cite{ack1}, Sec. 2.2.

There is, however,  one case where we generically reach the regular part of the envelope, as described by the following result.
\begin{theorem}[dS or AdS brane: Single bulk fluid with $\gamma\geq -1/2$]
In the above setup,  there are two possible, nonsingular  asymptotic behaviours corresponding to general (3 arbitrary constants) solutions of the field equations, and having the following properties:
\begin{itemize}
\item  $\gamma>-1/2$
\bq
\label{0_B2x}
x&=& \a\Upsilon+c_{-1\,1}-A\a/3c_{-2\,3}\Upsilon^{-1}+\cdots,\\
y&=& \a+A\a/3 c_{-2\,3}\Upsilon^{-2}+\cdots,\\
\label{0_B2w} w&=&  c_{-2\,3}\Upsilon^{-4}+\cdots,
\eq
where $c_{-1\,1}$ and $c_{-2\,3}$ are arbitrary constants. For $\Upsilon\rightarrow \infty$
we see that this is on $\Sigma_2$ given by Eq. (\ref{pfenv2}).

\item $\gamma=-1/2$
\bq
\label{-1/2_B3x}
x&=& \a\Upsilon+c_{-1\,1} \cdots,\\
y&=& \a \cdots,\\
\label{-1/2_B3w} w&=&  c_{-2\,3}\Upsilon^{-2}+\cdots,
\eq
where $c_{-1\,1} $ and $c_{-2\,3}$ are arbitrary constants.
Taking $\Upsilon\rightarrow \infty$ demonstrates that this is on $\Sigma_2$ given by Eq. (\ref{pfenv2}).
\end{itemize}
\end{theorem}
This result has two parts. The asymptotic behaviour was found in Ref. \cite{ack1}, Sec. 3.5. The envelope was derived in Ref. \cite{ack3}, App. A. We see that these universes look emptier at long distances into the bulk.

Further, there are curved brane solutions with regular support in the bulk coming not from the enveloping set but from the general solution of the field equations. The following result is shown in complete detail in Ref. \cite{ack4}.
\begin{theorem}No collapse singularity can arise in any brane model that comprises either
  \begin{itemize}
  \item a de Sitter brane in a single bulk fluid with negative energy density and $\gamma>-1/2$, or

  \item an Anti de Sitter brane in a single bulk fluid with positive energy density and $\gamma\in(-1,-1/2)$,
   \end{itemize}
      as the bulk scale factor is bounded from below and never vanishes.
\end{theorem}
We note that this result does not exclude the possibility of a big rip singularity in a finite distance from the brane position where the scale factor diverges.

However, whether or not such solutions are singular one may arrange for those solutions that allow for a jump discontinuity on the first derivative of the scale factor across the brane and satisfy the null energy condition, to \emph{match } for certain ranges of $\gamma$ producing non-singular universes \cite{ack4}. Unfortunately, the $\gamma$-ranges for the existence of such universes do not quite match other conditions on the fluid to localize gravity on the brane by requiring finiteness of the Planck mass there. This problem lies in the frontier of the singularity problem for such models.

There are a host of other regular asymptotic solutions describing  a curved brane sitting in a bulk with a coexisting, even slightly interacting,  dilaton-fluid system, cf. \cite{ack2}. It is an open problem to identify the structure of the enveloping brane in all these solutions, and so we do not discuss them any further here.

Another problem that is beyond our present results is what types of global bulk geometry are compatible with the asymptotic forms discussed here. If we suppose that away from the singularity the bulk space metric $g'_5$ is a kind a perturbation of the metric form assumed here, for instance \be g'_5=g_5+\delta g_5,\ee with $g_5$ given by Eq. (\ref{warpmetric}), then  what are the types of geometries which tend to the present ones discussed in previous Sections? This is somewhat reminiscent to the isotropization problem in inflationary cosmology.

Finally, we briefly comment about a different approach to the singularities in general brane-bulk systems. There is a basic issue of principle involved  in any discussion of singularities in the geometric context of a braneworld embedded in extra dimensions,  because the singularities present in the bulk away from the position of the brane, whose existence and nature we discussed in this paper, appear to be totally disconnected to the standard spacetime singularities predicted by the standard singularity theorems for the general relativistic metrics $g_4$. The same unconnectedness holds also for the cosmic censorship hypothesis  (presumably valid on the brane) which, in a truly higher-dimensional theory, ought to be perhaps an emerging property from structures which do not have a 4-dimensional counterpart. A way to connect the two is to extend the brane-bulk geometry in such a way as to allow our universe to be the conformal infinity of a certain 5-dimensional geometry, the \emph{ambient cosmological metric}. One then finds that the existence of the 4-dimensional spacetime singularities is constrained by the long-term, asymptotic properties of the ambient cosmological metric, while cosmic censorship holds true provided ambient space remains non-degenerate. For more details of this theory, we invite the reader to consult the recent review \cite{ac15} (which also includes the original papers).

\end{document}